\documentclass[preprint,showpacs,aps, preprintnumbers,amsmath,amssymb,superscriptaddress]{revtex4}

\usepackage{graphicx,graphics}   

\usepackage[utf8]{inputenc}
\usepackage{graphicx,amsfonts}
\usepackage{epstopdf}
\usepackage{subfig}
\usepackage{cancel,xcolor}
\usepackage{float}
\usepackage[english]{babel}
\usepackage{dcolumn}
\usepackage{bm}
\begin{document}

\title{ Zee and Zee-Babu mechanisms in the minimal 331 model}

\author{A. C. B. Machado}%
\email{a.c.b.machado1@gmail.com}
\affiliation{Centro de Ci\^{e}ncias Naturais e Humanas \\ Universidade Federal do ABC,
09210-580 Santo Andr\'{e}-SP, Brasil
}

\author{Pedro Pasquini}
\email{pasquini@ifi.unicamp.br}
\affiliation{~Instituto de F\'isica Gleb Wataghin - UNICAMP, {13083-859}, Campinas SP, Brazil}
\affiliation{Theoretical Physics Department, Fermi National Accelerator Laboratory, P. O. Box 500, Batavia, IL 60510, USA}

\author{V. Pleitez}%
\email{vicente@ift.unesp.br}
\affiliation{
    Instituto  de F\'\i sica Te\'orica--Universidade Estadual Paulista (UNESP)\\
    R. Dr. Bento Teobaldo Ferraz 271, Barra Funda\\ S\~ao Paulo - SP, 01140-070,
    Brazil
}

\date{10/04/2018}
%
\begin{abstract}
We show that the minimal 3-3-1 model cannot accommodate the neutrino masses at tree level using present experimental data. Nevertheless, a modified Zee and the Zee-Babu mechanisms for generating neutrino masses at 1-loop and 2-loop, respectively, are automatically implemented in the minimal 3-3-1 model, without introducing new degrees of freedom to the model. We also present a systematic method for finding solutions to the leptonic sector masses and mixing. As a case study, we accommodate the charged and neutral leptons masses and the PMNS matrix in the 1-loop modified Zee mechanism contained in the minimal 3-3-1 model.
\end{abstract}

\pacs{14.60.Lm,    
    14.60.Pq, 
12.60.Cn 
}

\maketitle

\section{Introduction}
\label{sec:intro}

At least two neutrinos are massive particles and present data provides reasonable knowledge to the leptonic mixing matrix, the so-called Pontecorvo-Maki-Nakagawa-Sakata (PMNS) matrix~\cite{Tanabashi:2018oca}. Since neutrinos are massive neutral fermions, they could be either Majorana or Dirac particles. At present many different mechanisms for generating neutrino masses are known but most of them are implemented making \textit{ad hoc} modifications of the particle content of a given model. 

For instance, the type I seesaw mechanism~\cite{Minkowski:1977sc}, includes both types of mass terms Dirac and Majorana. In these cases the mass eigenstates are Majorana fields and, the smallness of the neutrino masses is naturally explained. However, in this case, the PMNS matrix is just only approximately unitary and, if future data confirm that this matrix is unitary within the experimental error, the coexistence of both mass term will be ruled out. 

Things are different if neutrinos are strictly Dirac or Majorana fields: The PMNS matrix may be exactly unitary. Strictly Dirac neutrinos are obtained in left-right symmetric models~\cite{Senjanovic:1978ev} and strictly Majorana neutrinos are obtained in type II seesaw mechanism~\cite{Konetschny:1977bn} or also in left-right symmetric model with triplet scalars~\cite{Mohapatra:1979ia}.

Other interesting mechanisms that generate neutrino masses at 1-loop level~\cite{Cheng:1980qt,Magg:1980ut,Zee:1980ai} 
and 2-loop level~\cite{Cheng:1980qt,Zee:1985id,Babu:1988ki} and neutrinos are also strictly Majorana fields. In the latter cases, the smallness of the neutrino masses is explained because they are generated at the loop level.
These mechanisms are usually implemented as simple extensions of the standard model (SM) by \textit{ad hoc} inclusion of new fields aimed for generating the neutrino masses and are not naturally accommodated in the models.

This is particularly the situation in the mechanism in Refs.~\cite{Zee:1980ai,Babu:1988ki,Nomura:2018vfz}. Nevertheless, it is interesting to study the implementation of such mechanisms into self-contained models by the use of its particle content proposed for other theoretical reasons. Firstly because realistic models may contain not only one but several mass mechanisms within it. And secondly, because it is not obvious the possible correlations that may arise between neutrino mass mixing, matrices, and the other physical parameters. Hence, these are interesting benchmark scenarios to look for a rationale in physics beyond the standard model. That is, we can look at which models, proposed for other reasons, these mechanisms are implemented without introducing new particles to the model.

Here we will show that in the minimal 3-3-1 model (m331 for short)~\cite{Pisano:1991ee} neutrino masses can be implemented at tree level, but cannot be accommodated in present data. Nevertheless, the Zee and Zee-Babu mechanisms are natural consequences of the representation content of the model and we will show that these can accommodate all the leptonic sector masses and mixing by presenting a systematic method of finding solutions of Majorana neutrino masses and mixing.

\section{Leptons and scalars in the m331}
\label{sec:model} 
 
In the m331 model, the three lepton generations are all in triplets $\Psi_{aL}=(\nu_a\, l_a \,l^c_a)^T_L\sim(\textbf{3},0)$, where $a,l_a=e,\mu,\tau$.  In the scalar sector we have three triplets: $\eta=(\eta^0\,-\eta^{-}_1\,\eta^+_2)^T\sim({\bf3},0)$,
$\rho=(\rho^+\,\rho^0\,\rho^{++})^T\sim({\bf3},1)$,
$\chi=(\chi^-\,\chi^{--}\,\chi^0)^T\sim({\bf3},-1)$. Only the triplet $\eta$ and the anti-sextet~$S$
\begin{equation}
S=\left(
\begin{array}{ccc}
\sqrt{2}s^0_1& -s^+_1 & \frac{s^-_2}{\sqrt2}\\
-s^+_1& -\sqrt{2}S^{++}_1&\frac{s^0_2}{\sqrt2}\\
\frac{s^-_2}{\sqrt2}&\frac{s^0_2}{\sqrt2}&S^{--}_2
\end{array}
\right),
\label{sextet1}
\end{equation}
couple to the leptons through the Yukawa interactions $\overline{(\Psi_L)^c}G^S\Psi_LS$ and $\overline{(\Psi_L)^c}G^\eta\Psi_L\eta$, with $G^S(G^\eta)$ a symmetric (antisymmetric) matrix in the flavor space. 

Under the $SU(2)\otimes U(1)_Y$ group the scalar fields transform as
$\Phi_\eta=(\eta^0\, -\eta^-_1)^T$,
$\Phi_\rho=(\rho^+\, \rho^0)^T$, 
$\Phi_\chi=(\chi^-\, \chi^{--})^T$,
and $\Phi_s=
(s^-_2\, s^{0*}_2)^T$,
where these are doublets with weak hypercharge $Y=-1,+1,-3,+1$,  and the triplet 
\begin{equation}
T=i\tau_2\vec{\tau}\cdot\vec{t}=\left( \begin{array}{cc}
\sqrt{2}s^0_1 & -s^+_1 \\
-s^+_1 & -\sqrt{2}S^{++}_1
\end{array}
\right),
\label{triplet}
\end{equation}
is a triplet with $Y=2$. The $SU(2)$ singlets $\eta^+_2,\rho^{++},\chi^0,S^{--}_2$ have $Y=+2,+4,0,+4$, respectively. 

The total lepton number assignment in the scalar sector is
\begin{equation} 
L(T^*,\eta^-_2,\Phi_\chi,\rho^{--},S^{--}_2)=+2,\quad L(\Phi_{\eta,\rho,s},\chi^0)=0.
\label{ln}
\end{equation}
Notice that the only scalar doublet carrying lepton number is $\Phi_\chi$ and both members of the doublet have electric charge, for this reason, always $\langle \Phi_\chi\rangle=0$. The existence of scalars carrying lepton number implies the possibility of explicit breaking of this quantum number in the scalar potential. 

In the lepton sector the Yukawa interactions at the $SU(3)\otimes U(1)$ level are given by
\begin{equation}
-\mathcal{L}^{leptons}_Y=-\frac{1}{2}\epsilon_{ijk}\,\overline{(\Psi_{ia})^c}G^\eta_{ab} \Psi_{jb}\eta_k+\overline{(\Psi_{ia})^c}G^s_{ab} \Psi_{jb}S_{ij}+H.c.,
\label{yuka1}
\end{equation}
where $a,b$ are generations indices and $G^\eta$ ($G^S$) is an anti-symmetric (symmetric) matrix. Defining the multiplet according to the $SU(2)_{I,U,V}\otimes U(1)_{I,U,V}$, subgroups of $SU(3)\times U(1)_X$, transformations
\begin{equation}
L_a=\left(
\begin{array}{l}
\nu_{aL}\\ l_{aL}
\end{array}
\right)\sim(\textbf{2}_I,-1),\quad \Omega_{aL}=\left(
\begin{array}{l}
l_{aL}\\ (l^c)_{aL}
\end{array}
\right)\sim(\textbf{2}^*_U,0),\quad 
\Sigma_{aL}=\left(
\begin{array}{l}
\nu_{aL}\\ (l^c)_{aL}
\end{array}
\right)\sim(\textbf{2}^*_V,2)
\label{leptons1}
\end{equation}
with the definition $Q=I_3+Y/2$, $Q_U=2\bar{U}_3$ and $Q_V=\bar{V}_3+Y/2$.

The interactions in (\ref{yuka1}) can be written in terms of the $SU(2)_L\otimes U(1)_Y$ quantum numbers of the triplets and the sextet:
\begin{eqnarray}
2\mathcal{L}^l_Y&=&  -\overline{(\Sigma_L)^c_{la}}G^\eta_{ab}\epsilon_{pq}(\Sigma_L)_{kb}\eta^-_1+\overline{(\Omega_L)^c_{la}}G^\eta_{ab}\epsilon_{lk}(\Omega_L)_{kb}\eta^0+\bar{l}_{aR}G^S_{ab}\Phi^T_sL_b
\nonumber \\ &+& \overline{L^c_{ia}}G^S_{ab}
T_{ij} L_{jb}+ \overline{(L)^c_{ia}}\epsilon_{ij}G^\eta_{ab}L_{bj}\eta^+_2+\bar{l}_{aR}G^S_{ab}(l^c)_{bL}S^{--}_2 +H.c. 
\label{yuka2}
\end{eqnarray}

The full scalar potential has been consider in Ref.~\cite{DeConto:2015eia}, here we only include the $L$ violating interaction: 
\begin{equation}
f_3\left[\Phi^T_\eta T\Phi_\eta+
\frac{1}{\sqrt2}(\Phi^T_\eta\Phi_s+\Phi^T_s\Phi_\eta)\eta^+_2+ \eta^+_2\eta^+_2S^{--}_2\right]+f_4s^{0}_2s^-_2s^+_1+H.c.,
\label{21potential}
\end{equation}
where $f_3$ have dimension of mass. The interactions proportional to $f_3$ and $f_4$ come from the trilinear interactions $f_3\eta^TS^*\eta$ and 
$f_4\epsilon_{ijk}\epsilon_{mnl} S^*_{im}S^*_{jn}S^*_{kl}/3!$, respectively. It is clear from the Yukawa interaction with $\eta^+_2$ in (\ref{yuka2}), and the trilinear interaction $f_3\eta^0s^-_2\eta^+_2$, that the Zee's mechanism~\cite{Zee:1980ai} for generating neutrino masses at 1-loop is automatically implemented in this model. Notice that the implementation is not the original Zee mechanism, since there are also another contribution to the 1-loop mechanism coming from the interaction with the triplet in (\ref{yuka2}) $f_3\Phi^T_\eta T\Phi_\eta$ given by the trilinear $f_3\eta^0s^+_1\eta^-_1$. The term with $f_4$ in Eq.~(\ref{21potential}) involves the interactions of the doublet $\Phi_s$ and the triplet $T$ and will generate a 1-loop diagram involving the symmetrix matrix $G^S$.
Moreover, the interaction in Eq.~(\ref{yuka2}) with the doubly charged $S^{--}_2$ and the trilinear $f_3\eta^+_2\eta^+_2S^{--}_2$ imply that the 2-loop mechanism in \cite{Babu:1988ki} is also implemented.

We write here only the constraint equation for $v_{s_1}$. It reads~\cite{DeConto:2015eia}:
\begin{equation}
t_{s_1}=v_{s_1}[2(e_1+e_2)v^2_{s_1}+2e_1v^2_{s_2}+(c_1+d_3+d_4)v^2_\eta+d_5v^2_\rho+d_1v^2_\chi+\mu^2_S]+\frac{1}{\sqrt2}f_3v^2_\eta-2\sqrt{2}f_4v^2_{s_2},
\label{vs1}
\end{equation}
from which wee see that $t_{s_1}=0$ implies $v_{s_1}=0$ only if $f_3=f_4=0$. This is not the case here and we see that allowing $L$ violating terms implies a non-vanishing $v_{s_1}$. In fact, the diagrams inducing a neutrino mass at 1- and 2- loop, imply divergent contribution to the tadpole $v_{s_1}$. Hence a counter-term is necessary to be added in the Lagrangian which implies that $v_{s_1}\not=0$~\cite{Pleitez:1993gc}. However, we want to have a rather small $v_{s_1}$ i.e., one which does not give a relevant contribution to the neutrino masses. It means that we can impose $(v_{s_1}+\delta^n v_{s_1})/v_{s_2}\ll1$ at the $n$-loop order.   
Anyway, $v_{s_1}$ is small at the tree level: $v_{s_1}\approx -(f_3v_\eta/d_1v^2_\chi)v_\eta$, if $f_3<0$, but $\vert f_3\vert\gg f_4$ or, $v_{s_1}\approx 2\sqrt{2}(f_4v_{s_2}/d_1v^2_\chi)v_{s_2}$ if $f_4\gg \vert f_3\vert$ , and $0<d_1\approx1$. 

Notice that, since the $SU(2)$ doublets carry a label a\-ccor\-ding to the triplet of $SU(3)$ to which they belong, it is not possible to choose a different weak basis~\cite{DeConto:2015eia}.

We see that, at the level of $SU(2)_L\otimes U(1)_Y$, the antisymmetrical character of the matrix $G^\eta_{ab}=-G^\eta_{ba}$ and the symmetric of $G^S_{ab}$(=$G^S_{ba}$) have not to be imposed by hand, and the Zee's mechanism with a singly charged scalar~\cite{Zee:1980ai} and the triplet of the type-II seesaw mechanism~\cite{Konetschny:1977bn} appear naturally in this model since there are two Higgs doublets which couple to leptons, $\Phi_\eta$ and $\Phi_s$ and the singly charged singlet $\eta^-_2$. The Zee-Babu mechanism is also implemented by the doublets, the singlets $\eta^-_2$ and $S^{--}_2$ with the trilinear in Eq.~(\ref{21potential}).  

\section{Lepton masses at tree level}
\label{sec:tree}

Using the representation in Eq.~(\ref{leptons1}) in Eq.~(\ref{yuka2}) the lepton mass matrices are given by
\begin{equation}
M ^\nu=\frac{v_{s_1}}{\sqrt2}G^S_{ab},\quad
M^l_{ab}=\frac{v_\eta}{\sqrt2}G^\eta_{ab}+\frac{v_{s_2}}{2}G^S_{ab}.
\label{leptons2}
\end{equation}

Active neutrinos are Majorana fields and there is flavor changing neutral interactions in the scalar sector.  Notice that if $v_{s_1}=0$  at tree level the neutrino are massless and have no connection to the charged lepton masses. However, this is not possible if $L$ violating terms do exist in the scalar potential inducing a tadpole diagram at the 1-loop level since it is necessary to add a counterterm $\langle s^0_1\rangle\not=0$ even at three level~\cite{Pleitez:1993gc}. 

We will show that at tree level even if $\langle s^0_1\rangle\not=0$ it is not possible from Eqs.~(\ref{leptons2}) to fit the observed charged lepton and neutrino masses at the same time.

\begin{table}[!ht]
\centering
\begin{tabular}{cccc}\hline
parameter  & value & error\\ \hline \hline
$\Delta m_{21}^2/10^{-5}$ & 7.56 $\rm eV^2$ & (19)  \\ 
$\Delta m_{31}^2/10^{-3}$ & 2.55 $\rm eV^2$ & (4) \\ 
$\sin^2\theta_{12}$       & 0.321           & (18) \\ 
$\sin^2\theta_{13}$       & 0.02155         & (90)\\ 
$\sin^2\theta_{23}$       & 0.430           & (20) \\    
$\delta_{\rm CP}/\pi$     & 1.40            & (31) \\ 
$m_e$     & 0.5109989461 MeV       & (31) \\ 
$m_\mu$     & 105.6583745 MeV          & (24) \\ 
$m_\tau$     & 1776.86   MeV          & (12) \\ \hline
\end{tabular}
\caption{\label{tab:3nubest} Current best fit values of Standard-3$\nu$ as given by~ \cite{deSalas:2017kay} and~\cite{Patrignani:2016xqp}.}
\end{table}

From Eq.~(\ref{leptons2}) we can write 
\begin{equation}
G^S=\frac{\sqrt{2}}{v_{s_1}}M^\nu,\quad M^l=\frac{v_\eta}{\sqrt{2}} G^ \eta+\frac{v_{s_2}}{v_{s_1}} M^\nu,
\label{zb0}
\end{equation}
Notice that we are assuming that $v_{s_1}$ is different from zero, i.e., that there is a tree level contribution to the neutrino masses. Notice also that since the charged lepton masses have two contributions, the model predicts flavor changing neutral currents in the lepton sector through neutral scalars. If $G^\eta=0$ both $M^\nu$ and $M^l$ are proportional to the matrix $G^S$ and the PMNS matrix is just th unit matrix.

For shortness, we will write $\overline{M}^a=M^a.(M^a)^\dagger$, $a=l,\nu$. In any basis, the invariants of $\overline{M}^a$ are,
\begin{eqnarray}
I_1^a={\rm Tr}[\overline{M}^a]=m^2_1+m^2_2+m^2_3, \nonumber \\  
I_2^a={\rm Tr}[(\overline{M}^a)^2]=m^4_1+m^4_2+m^4_3, \nonumber \\ 
I_3^a={\rm Det}[\overline{M}^a]=m^2_1 m^2_2 m^2_3
\label{invariants}
\end{eqnarray}
where $m_i$ are the masses of the particles described by matrix $M^a$. Those equations can be used to find all the eigen-values and all other invariantes (for example, ${\rm Tr}[(\overline{M})^3]=\frac{1}{2}(I_1 I_2-I_1^3)+I_3$) and must be obeyed by all the free parameters.
Now, using the $I^{\nu,l}_1$ invariants defined in Eq.~(\ref{invariants}) and the matrix $M^l$ in Eq.~(\ref{zb0}) the first invariant requires that,

\begin{equation}
v^2_\eta (\vert G^\eta_{12}\vert^2+\vert G^\eta_{13}\vert^2+\vert G^\eta_{23}\vert^2) + \left(\frac{v_{s_2}}{v_{s_1}}\right)^2 I_1^\nu= I_1^l,
\label{inv2}
\end{equation}
where the first term on the left side arises from $\textrm{Tr}G^\eta G^{\eta\dagger}$.
This equation tightly bounds the maximum values of all the modulus of the free parameters to be no greater than $\sqrt{I^l_1}$. In special, we can think of them as a 4-dimensional sphere of radious $R$ described by a four dimensional vector on an Euclidian space of the form,
\begin{equation}
\vec{x}=(v_\eta G^{\eta}_{12},v_\eta  G^{\eta}_{13},v_\eta G^{\eta}_{23}, \frac{v_{s_2}}{v_{s_1}} \sqrt{I_1^\nu})=R(e^{i\delta_1} s_\phi s_\theta c_\omega,e^{i\delta_2} c_\phi s_\theta c_\omega, e^{i\delta_3} c_\theta c_\omega,s_\omega),
\label{4sphere_1}
\end{equation}
with $c_x=\cos x$ and $s_x=\sin x$ and $\delta_i$ are the complex phases of the parameters $G^\eta_{ij}$. Thus, we can write Eq.~(\ref{inv2}) as
\begin{equation}
|\vec{x}|^2=R^2=I_1^l,
\end{equation}
Eq.~(\ref{4sphere_1}) allows us to find the relevant parameters that constrains the charged lepton masses in therms of the radius $R$ and the angles $\phi,\theta$ and $\omega$. This parametrization is useful in many ways. Firstly, it straightforwardly solves the constraint of the invariant $I_1^l$ by defining one single mass scale $R$. Secondly, the $\omega$ parameter is defined in such a way that it regulates the relative contribution of each matrix, $G^\eta$ and $M^\nu$(or $\sqrt{I^\nu_1}$) to the charged lepton mass:
\begin{equation} 
G^\eta\propto \cos \omega,\quad 
M^\nu\propto \sin \omega.
\end{equation}
That means a $\omega\rightarrow0$ corresponds to the lepton masses completely anti-symmetric ($G^\eta$ dominated) or $\frac{v_{s_2}}{v_{s_1}} M^\nu<<\frac{v_\eta}{\sqrt{2}}G^\eta$ while $\omega\rightarrow \pi/2$ is the other extreme ($M^\nu$ dominates) where the charged lepton matrix and the neutrino matrix are proportional to each other, $\frac{v_\eta}{\sqrt{2}}G^\eta<<\frac{v_{s_2}}{v_{s_1}} M^\nu$.

\begin{figure}[H]
    \centering
    \includegraphics[scale=0.4]{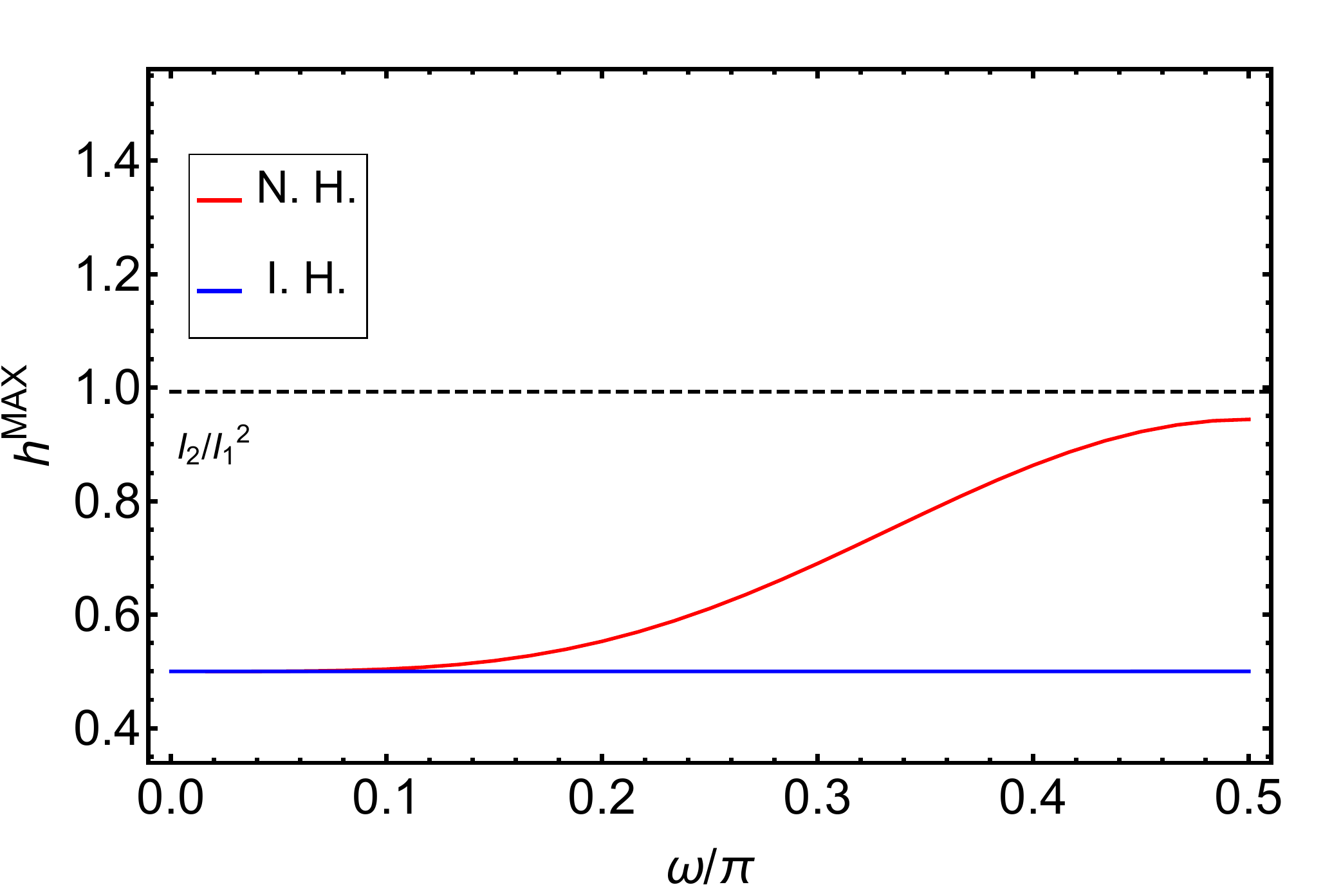}
    \caption{\label{fig:f_2} Possible maximum values of the $h$ function by fixing the $\omega$ variable. The Red line is assuming Normal Hierarchy, blue line is Inverted Hierarchy and black-dashed line is the fraction $I_2^l/(I_1^l)^2$.}
\end{figure}

Now, using the parametrization in Eq,~(\ref{4sphere_1}), that is $v_\eta G^\eta_{12}\to Re^{i\delta_1}s_\phi s_\theta c_w$, etc and $G^S_{ij}=\sin\omega \hat{m}_i\delta_{ij}$ with $\hat{m}_i=m^\nu_i/\sqrt{I^\nu_1}$ in the invariants $I^l_2$ given in Eq.~(\ref{invariants}) we obtain: 
\begin{equation}
(I_1^l)^2 h\left(\theta, \phi,\omega, m_i^2/I_1^\nu,\delta_i\right)=I_2^l,
\label{yuca1}
\end{equation}
where $h$ is given by 
\begin{eqnarray}
h&=&s^4_\omega\frac{I^\nu_2}{(I^\nu_1)^2}+\frac{1}{2}c^4_\omega-s^2_\omega c^2_\omega\left\{ s^2_\theta[\hat{m}^2_1+(\hat{m}^2_3-\hat{m}^2_2)c_{2\phi}+ 2c^2_\theta(\hat{m}^2_3+\hat{m}^2_2)]\right.\nonumber \\&+&\left.
2\hat{m}_1s^2_\theta(\hat{m}_2s^2_\phi c_{2\delta_1}+2\hat{m}_3c^2_\phi c_{2\delta_2})+\hat{m}_2\hat{m}_3c^2_\theta c_{2\delta_3}\right\}
\label{h}
\end{eqnarray}

i.e., it depends on the neutrino mass square, the three angles, and three phases. From Eq.~(\ref{yuca1}) we see that it is possible to fit the charged lepton and neutrino masses at the same time only if $h=I_2^l/(I_1^l)^2$. In fact, inserting current values of lepton masses we get $I_2^l/(I_1^l)^2\approx 1$, because lepton masses are very hierarchical. On Fig.~\ref{fig:f_2} we show the maximum value of $h$ by fixing $\omega$ for Normal Hierarchy (red sinuous line) and Inverted Hierarchy (blue bottom line), the Black-Dashed upper line corresponds to the value that fits the invariant $I_2^l$. 

As we said above, the function $h$ that correctly fits the charged lepton masses corresponds to the intersection point between a colored curve and the black-dashed line. The blue bottom curve is constant with value $0.5h^{\textrm{max}}$ and the red curve approaches the line but has its maximum value at $\omega/\pi=0.5$ below the dashed line, even if one takes into account lepton and neutrino mass errors. Thus, Eq.~(\ref{leptons2}) cannot provide a solution to the parameters of Table~\ref{tab:3nubest}. This happens because the Eigenvalues of $M^\nu$ and $M^l$ are related, see for example Eq.~(\ref{inv2}). This means that the hierarchy between the neutrino masses controls the maximum hierarchy reachable by the lepton masses. As charged lepton masses are more hierarchical than the neutrino masses, $\Delta m_{\tau \mu}^2/\Delta m_{\tau e}^2=0.99646$ while $\Delta m_{21}^2/|\Delta m_{31}^2|=0.0296$, where $\Delta m_{ab}^2=m_a^2-m_b^2$, that means we cannot fit both at the same time. In order to find such solution, one needs $I_2^\nu/(I_1^\nu)^2\geq I_2^l/(I_1^l)^2$, which cannot be accomplished with the values presented in Table~\ref{tab:3nubest}. Notice that
\begin{enumerate}
\item If $\omega=0$, or $G^S\ll G_\eta$, $h(\omega=0)=1/2$.
\item $\omega=\pi/2$, or $G^\eta \ll G^S$, $h(\omega=\pi/2)=I^\nu_2/(I^\nu_1)^2$. 
\item $h(\omega=0)\leq h(\omega)\leq h(\omega=\pi/2)$, it means that we can fit $I^l_2$ only if 
\begin{equation}
\frac{1}{2}\leq \frac{I^l_2}{(I^l_1)^2}\leq  \frac{I^\nu_2}{(I^\nu_1)^2} ,
\label{touche}
\end{equation} 
which is not compatible with the data.
\end{enumerate}

Hence, we have shown that in the m331 model, neutrinos cannot obtain realistic masses at tree level. Loop corrections have to be taken into account. 

\section{Neutrino masses at one and two loops}
\label{sec:nusmass}

Having demonstrated that at tree level the model can not generate the correct charged lepton masses this also implies that the leptonic mixing PMNS matrix cannot be obtained. A possibility is to add sterile (under the 3-3-1 symmetry) right-handed neutrinos and implement the type I seesaw as in Ref.~\cite{Machado:2016jzb}.
However, one can wonder if the own model content provides a solution to this issue. The answer is yes if not only the tree level contribution is taken into account, but also 1- and/or 2-loop corrections, through the Zee and Zee-Babu mechanism, respectively, which are both naturally implemented with the minimal representation content of the model, it means without introducing new degrees of freedom. As can be seen from Eq.~(\ref{yuka2}) the model has the charged scalar singlet of $SU(2)_L$,  $\eta^+_2$ as in the Zee model~\cite{Zee:1980ai} and also the doubly charged singlet scalar $S^{--}_2$ of the 2-loop mechanism  of Refs.~\cite{Cheng:1980qt,Zee:1985id,Babu:1988ki}. Since the model also has the respective trilinear interactions in Eq.~(\ref{21potential}), both mechanisms for generating neutrino masses are naturally implemented in this model. However, we stress that it is necessary to have a tree level contribution too.

\subsection{Neutrino masses with 1-loop contributions}
\label{subsec:zee}
We first start by a defining the normalized the mass matrices as follows: 
\begin{eqnarray} 
&&\hat{M}^l=\frac{M^l}{\sqrt{I_1^l}}, \qquad {\hat M^\nu}= \frac{M^\nu}{\sqrt{I_1^\nu}}, \qquad
\hat{G}^S=\frac{v_{s_2}}{\sqrt{2}}\frac{G^S}{\sqrt{I_1^l}},\qquad  {\hat G^\eta}=\frac{v_\eta}{\sqrt{2}}\frac{G^\eta}{\sqrt{I_1^l}}
\label{norma}
\end{eqnarray}
defining $I_1^a$ as in Eq.~(\ref{invariants}). This is usefull because we can get rid of mass scale and treat all the parameters as adimensional.

Using the definitions in Eq.~(\ref{norma}), the tree level charged lepton masses and the tree level plus a 1-loop contributions to the neutrino masses are written: 
\begin{eqnarray}
\hat{M}^l&=&{\hat G}^\eta+{\hat G}^S, 
 \nonumber \\ 
\hat{M}^\nu&=&a_0{\hat G}^s+a_1 {\hat G}^s({\hat G}^S)^\dagger{\hat G}^s+a_2 \left[{\hat G}^\eta({\hat G}^\eta+{\hat G}^S)^\dagger{\hat G}^S+{\rm Transpose}\right]
\label{leptonmasses}
\end{eqnarray}
where $a_0$ denotes the tree level mass, and
\begin{eqnarray} 
a_0=\frac{v_{s_1}\sqrt{I_1^l}}{v_{s_2}\sqrt{I_1^\nu}},\quad 
a_1=a_0\left(\frac{4f_4 I_1^l}{m^2_{s^-_1}v_{s_2}}\right)\ln\left(\frac{m_{s^-_2}^2}{m_{s^-_2}^2}\right),\quad
a_2=a_0\left(\frac{2f_3 I_1^lv_{s_2}}{m^2_{\eta^-_2}v^2_{s_1}}\right)\ln\left(\frac{m_{\eta^-_2}^2}{m_{s^-_2}^2}\right),
\label{as}
\end{eqnarray}
In fact, we have verified that if $v_{s_1}=0$, i.e., no neutrino masses at tree level, the 1-loop only does not generated the correct masses for neutrinos. Hence, using the parametrization in Eq.~(\ref{as}) we have already assumed that $v_{s_1}\not=0$.

Notice that the free parameters are the VEVs $v_{s_1,s_2},f_3,f_4$ and the masses of the charged scalars, $m_{s^-_1},m_{s^-_2},m_{\eta^-_2}$. All of them will appear in other places of the Lagrangian but, for the leptonic masses only the ratios in $a_0,a_1$ and $a_2$ matters. Notice that the term with $a_1$ is the contribution of the triplet $T$ which does not exist in the Zee model, the latter model contains only one scalar with anti-symmetric coupling, this model has an extra scalar that couples symmetrically with the fermions see Eq.~(\ref{21potential}). 

Hence, given $a_0$, $a_1$ and $a_2$ will allow us to adjust both charged lepton and neutrino mass which creates an opportunity to find a solution. The existence of those new parameters implies that one can start with a general lepton matrix $\hat{M}^l$ constituting of a sum of an anti-symmetric matrix ${\hat G}^\eta$ and a symmetric matrix ${\hat G}^S$ and later adjust the neutrino mass matrices.


We show that in this case Eqs.~(\ref{leptonmasses}), which add the 1-loop correction to the tree level mass matrix, has a possible solution which accommodate the measured mixing parameters of Table~\ref{tab:3nubest} at less than 1-$\sigma$. Using $v_\eta=240$ GeV and $v_{s_2}=1$ GeV, and $\sqrt{I^l_1}=1.78$ GeV (using the central values of $m_l$), we obtain
\begin{eqnarray}\nonumber
G^\eta&=&\left(
\begin{array}{ccc}
0 & 1.750-0.23i & 1.755-0.383i\\
-1.750+0.23i &0 & 0.376-0.027i \\
-1.755+ 0.383i & -0.376+0.027i & 0
\end{array}
\right)\times 10^{-3}\nonumber \\
G^S&=&\left(
\begin{array}{ccc}
0.499+1.719i & 0.690+0.518i & 0.426+0.320i\\
 & 0.416 & 0.256\\
 &  & 0.158
\end{array}
\right)
\end{eqnarray}
with $a_0=0.202478$, $a_1=0.5$ and $a_2=1.90921$. 


Thus, the matrices that rotate the leptonic mass matrix and the neutrino mass matrix can be obtained and we write them explicitly,

\begin{align}\nonumber
 V^l_L=&\left(\begin{array}{ccc}
      -0.044+0.147 i & 0.282 -0.627 i & 0.0260 +0.709 i\\
      -0.286-0.134 i &    0.4614 +0.454 i & 0.664 +0.204 i \\
      0.786 -0.508 i & 0.315 -0.0988 i & 0.1049 -0.0621 i 
     \end{array}\right) \\
V^\nu=&\left(\begin{array}{ccc}
 -0.148+0.0576 i & -0.164-0.200 i & 0.822 -0.483 i \\
0.695 -0.110 i & 0.523 +0.386 i & 0.237 -0.160 i \\
-0.613+0.321 i & 0.672 +0.243 i & 0.00136 -0.101 i
     \end{array}\right)
     \label{massnum}
\end{align}

This predicts a normal hierarchy solution with $m_{\rm lightest}=0$, $\Delta m_{21}^2/10^{-5}=7.56 \,{\rm eV}^2$ and $\Delta m_{31}^2/10^{-3}=2.55\, {\rm eV}^2$. The mixing angles can be obtained by the equation,
\begin{equation}
U_{\rm PMNS}=V^{l^\dagger}_LV^\nu
\end{equation}
Notice that the numerical solution gives a $U_{\rm PMNS}$ containing all its possible phases, some of them are physical and some of them can be absorbed by a rotation in the charged leptonic sector. In the PDG~\cite{Tanabashi:2018oca} notation we can sepparate the PMNS matrix into three sub-matrices,
\begin{equation}
U_{\rm PMNS}=P_l^\dagger.V_{\rm PMNS}.P
\end{equation}
where $V_{\rm PMNS}$ contains the 3 mixing angles and the Dirac CP phase. $P$ contains the majorana phases and can be written as $P={\rm Diag} [1, e^{\alpha_{21} i/2},  e^{\alpha_{31} i/2}]$ while $P_l$ is a diagonal matrix with the unphysical phases that are absorbed by a redefinition of the cherged lepton basis. In this notation, we can identify the mixing angles as $\sin^2\theta_{12}=0.318$, $\sin^2\theta_{13}=0.02043$, $\sin^2\theta_{23}=0.373$, and $\delta_{\rm CP}=1.29\pi$, which are inside the $1\sigma$ range of Table~\ref{tab:3nubest}, and produces the matrix,
\begin{equation} 
V_{\rm PMNS}=\left(
\begin{array}{ccc}
0.8193 & 0.5552 & 0.14293 e^{-1.29\pi i}\\
0.4032 e^{3 i} &0.6868 e^{0.0560 i}& 0.60479 \\
0.4076e^{0.1812 i} & 0.4691e^{3.0353 i} & 0.7834
\end{array}
\right).
\label{pmns}
\end{equation}
Since the neutrinos are majorana, the majorana phases could be important in some physical process, thus, for completeness, we present them here: $\alpha_{21}=1.15\pi$ and $\alpha_{31}=0.862\pi$.

\subsection{2-loop neutrino masses}
\label{subsec:zeebabu}

The m331 model also allows one to give mass to neutrinos at 2-loop using the Babu mechanism. In this mechanism, worked in more details for Babu~\cite{Babu:1988ki}, it is necessary to have the Zee's singly charged scalar, here $\eta^+_2$, and a singlet doubly charged scalar, here $S^{--}_2$. Both are singlet of $SU(2)$. See the Yukawa interactions in Eq.~(\ref{yuka2}) 
and the scalar interactions in Eq.~(\ref{21potential}) with the loops obtained by two $\eta^+_2$ a one $S^{--}_2$, assuming $m^2_{S^{--}_2}>m^2_{\eta^-_2}$, it follows~\cite{Babu:1988ki}
\begin{equation}
M^\nu_{ab}\simeq \frac{
    G^\eta_{ac}G^S_{cd}
}{
    32\pi^2
}
\frac{
    m_cm_d
}{
    m^2_{S^{--}_2} 
}G^{\eta\dagger}_{db}
\left[\ln 
\left(\frac{
    m^2_{\eta^-_2}+m^2_{S^{--}_2}
}{
    m^2_{\eta^{--}_2}
}
\right)
\right]^2f_3,
\label{babuint}
\end{equation}
where $m_c,m_d$ are the charged lepton masses in the internal line. When obtaining (\ref{babuint}) it was assumed that $m_{S^{--}_2}>m_{\eta^-_2}$. Detalis of the contributions to the neutrino masses by the Zee-Babu mechanism will be shown elsewhere.

\section{Neutrino masses with 1-loop contributions}
\label{sec:solutions}

Now, we can redo the analysis of the previous section to show
that Eq.~(\ref{leptonmasses}) have solutions which are compatible with the observed values for the lepton masses. We will denote by a subscript 2 the basis where the matrix $G^S$ is diagonal (see the Appendix), this $G^S_2={\rm Diag}[g_1,g_2,g_3]$ where its eigenvalues are $g_i,\; i=1,2,3$. Since these are free parameters the matrix $G^S$ is not bounded by the neutrino masses as $M^\nu$. 

\begin{figure}[H]
\centering
\includegraphics[scale=0.4]{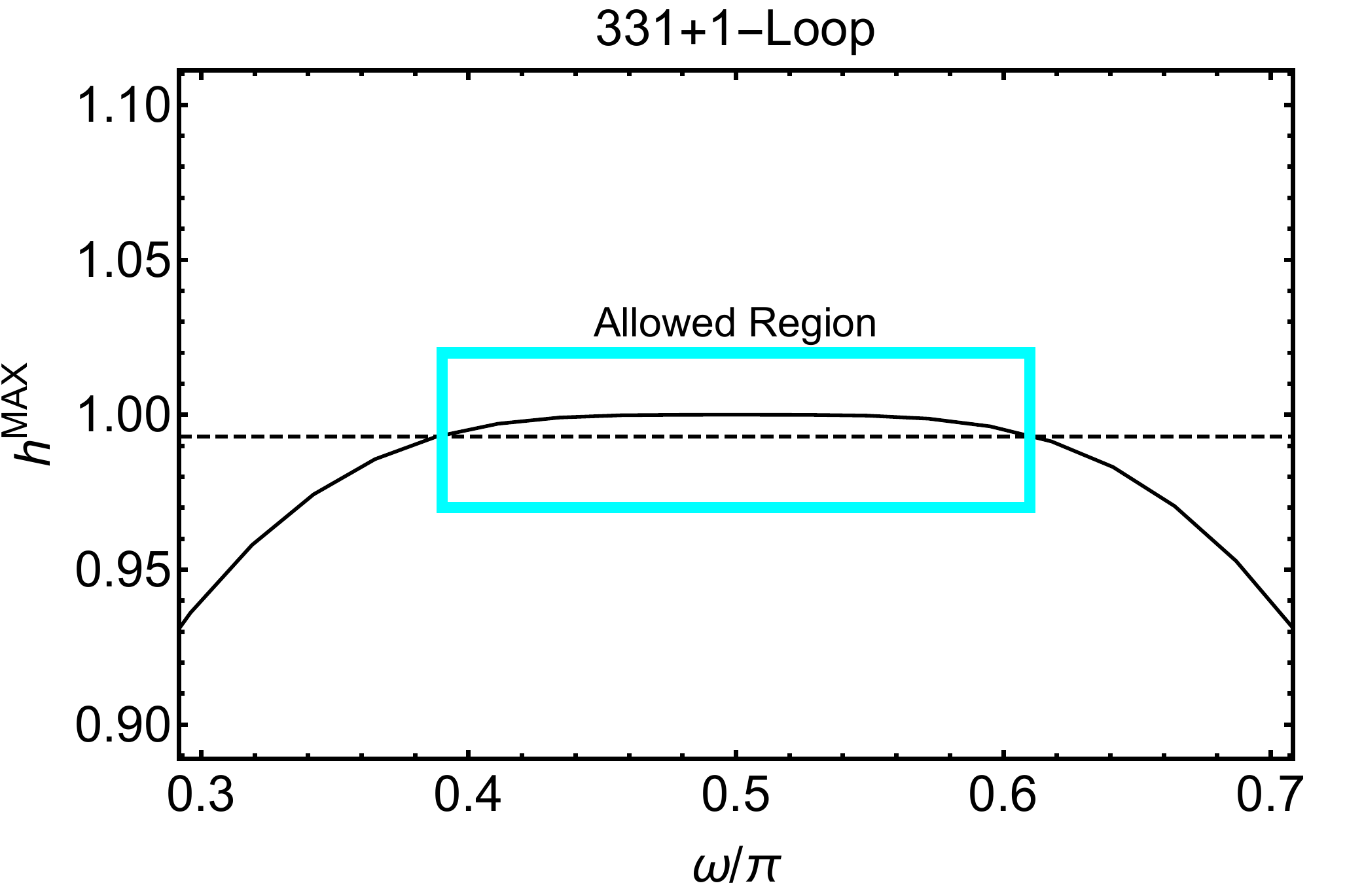}
\caption{\label{fig:f_2_1loop} Possible maximum values of the $h$ function by fixing the $\omega$ variable (black) and black-dashed line is the fraction $I_2^l/(I_1^l)^2$.}
\end{figure}

We will show that, with the same methodology as in the previous section, neutrino masses and the PMNS matrix can be accommodated as it was shown in Sec.~\ref{subsec:zee}. The first charged lepton mass invariant gives
\begin{equation}
2(|(\hat{G}^{\eta}_{2})_{12}|^2+|(\hat{G}^{\eta}_{2})_{13}|^2+|(\hat{G}^{\eta}_{2})_{23}|^2)+\sum_i g_i^2= 1
\end{equation}
This equation is equal to 1 because of the normalization ( $\textrm{Tr}(\hat{\bar{M}})=1$). We can now define a 4-dimensional sphere, but now its radius is $R=1$ and the four dimensional vector on an Euclidean space of the form,
\begin{equation}\label{eq:4sphere}
\vec{x}=(\sqrt{2}(\hat{G}^{\eta}_{2})_{12},\sqrt{2} (\hat{G}^{\eta}_{2})_{13},\sqrt{2} (\hat{G}^{\eta}_{2})_{23}, \sum_i g_i^2)=(e^{i\delta_1} s_\phi s_\theta c_\omega,e^{i\delta_2} c_\phi s_\theta c_\omega, e^{i\delta_3} c_\theta c_\omega,s_\omega)
\end{equation}
with $c_x=\cos x$ and $s_x=\sin x$ and $\delta_i$ are the complex phases of the parameters $\hat{G}_{\eta 2}^{ij}$. Now, $\omega$ controls the symmetry proportion of $\hat{M}^l$. If $\omega\rightarrow 0$ it is anti-symmetric and if $\omega\rightarrow \pi/2$ it is symmetric. Next, one needs to find a solution to the equation
\begin{equation}
I^l_2\equiv {\rm Tr}[(\hat{M}^l\hat{M}^{l\dagger})^2]=(I^l_1)^2h\left(\theta, \phi,\omega, \delta_i,\hat{g}_i^2\right).
\end{equation} 
We will not write $h$ explicitly but notice that now, since the parameters $g_i$ are free, we can find a solution for all letpton masses. The plot of the maximum possible values $h$ is presented on Fig~\ref{fig:f_2_1loop}.
Notice that this form of equation restrains the value of $\omega$ to $\omega\in[69^\circ,110^\circ]$. Where $\omega=\pi/2$ imply $G^\eta_2=0$.

Also, because of the high hierarchy of the lepton masses, $g_i$ will also be hierarchical. There are several solutions to this equations and one can solve the invariants $I_2^l$ and $I_3^l$ as a function of the variables $g_i$, $\omega, \phi,\theta$ and the $\delta_i$ phases. 

Going back to the 1-loop Eq.~(\ref{leptonmasses}) the strategy to find the general solution is:
\begin{itemize}
\item [(i)] Use one $g_i$ and two out of three angles  $\omega$, $\theta$ and $\phi$ to satisfy the other two charged lepton invariants $I_i^l$, $i=2,3$.
\item [(ii)] Use the free parameters $a_i$, $i=0,1,2$, $m_{\rm lightest}$ and two remaining $g_i$ to find the solutions to the neutrino invariants $I_i^\nu$, $i=1,2,3$.
\item [(iii)] Now one can write $G^\eta_2$ and $G^S_2$ in the auxiliary basis and find $U_2$ and $V_{L2}$.
\item [(iv)] Now go back to the general basis where $G^S$ is not diagonal by the use of a unitary matrix $A$, $G^S=A^T.G^S.A$. Now, the free parameters of $A$ and the previously found $U_2$ and $V_{L2}$ can be used to fit the PMNS matrix through Eq.~(\ref{eq:PMNS_2}). One can use the phases $\delta_i$ to better fit the parameters. Notice, however that only 1 of those are physical. Thus, one can set $\delta_2=\delta_3=0$. 
\end{itemize}

The particular solution found above was obtained by this method. In there we settled: $g_1=g_2=0$ and $\phi=\pi/4$. In this case, the invariants ${\rm Tr}[M^lM^{l\dagger})^2]$ and ${\rm Det}[M^lM^{l\dagger}]$ imply $\theta=1.41^\circ$ and $\omega=69.91^\circ$ and the lightest neutrino mass to be equal to zero.
This means that the method of finding charged lepton masses and mixing here described are general to any model that does not restrain the lepton matrix and we found a set of conditions that any charged lepton matrix should obey in order to fit the invariants $I_i^l,\;i=1,2,3$ defined in Eq.~(\ref{invariants}).


\section{Conclusions}
\label{sec:con}

Here we have shown that in the m331 model active neutrinos may be pure Majorana particles and the PMNS matrix truly unitary.
The Majorana mass term has three contributions: i) the tree level one induced by a non-zero VEV, $v_{s_1}$, the 1-loop contribution arisen from the Zee mechanism, and the 2-loop contributions by the Zee-Babu mechanism. Each of these contributions may be more or less important. We showed that the tree level-only cannot fit current neutrino oscillation data, but 1-loop can.

No right-handed (sterile under the 331 gauge symmetry) neutrinos are needed. This sort of neutrinos may be introduced just to explain the dark matter relic density or, if in the future 
right-handed neutrinos are found \cite{Aguilar-Arevalo:2018gpe,Almazan:2018wln} or, if the PMNS matrix becomes not exactly unitary, we will have to introduce right-handed neutrinos singlets of $SU(3)_L\otimes U(1)_X$. However, even in this case, an important part of the neutrino masses may arise from the mechanisms analyzed in this work. 
If this were the case, it is interesting that right-handed neutrinos do not need to be very heavy because they will not be needed to implement the seesaw type I mechanism. They could be light enough to explain any anomaly (dark matter?) if it existed and need to be solved by light right-handed neutrinos. 

The m331 model in intrinsically a multi-Higgs model and at least with the actual data, the existence of extra scalars, besides that observed one with a mass of about 125 GeV, cannot be excluded.
The m331 model has all the scalar multiplets that have been introduced by hand in the context of the SM model and all of them are needed to break the gauge symmetries and/or to generate all fermion masses. This includes several doublets, and neutral, singly charged and doubly charged singlet and triplet scalars. Notice that, as in any multi-Higgs model, the properties of the Higgs boson with a mass of 125 GeV are different from those in the SM model. Moreover, since the measurements of properties of the would-be the SM Higgs scalar do still allow the considerable non-standard behavior of the 125 Higgs boson~\cite{Lenz:2018cce}. The extra scalar fields have interesting phenomenological consequences, for instance $pp\to X^{++}X^{--}\to l^+l^+l^{'-}l^{'-}$, where $l,l'=e,\mu$ and $X^{++}$ may be a doubly charged scalar or vector field~\cite{Coriano:2018coq,Chang:2018wsw,Crivellin:2018ahj}.

\acknowledgments
VP thanks to CNPq for partial financial support. All authors are thankful for the support of FAPESP funding Grant No. 2014/19164-6. P. P. thanks the support of FAPESP-CAPES funding grant 2014/05133-1 and 2015/16809-9, also the partial support from FAEPEX funding grant, No 2391/17, Fermilab NPC fellowship and APS-SBF fellowship.

\appendix

\section{Systmatic for finding solutions fitting lepton masses and the PMNS matrix}
\label{sec:sol}

The two basis that will be relevant in order to present a systematic way of finding solutions are described below.
\begin{enumerate}
\item  {\bf General mass matrix :}\\
This is the general basis with both $\hat{M}^\nu$ and $\hat{M}^l$, defined in Eq.~(\ref{norma}), are as general as possible. To go from flavor eigenstates to the mass eigenstes, one can by an unitary matrix $V^\nu$ rotate $M^\nu$, and $M^l$ by two matrix $V^l_L$ and $V^l_R$: The matrices $\hat{M}^\nu$ $\hat{M}^l$ below are diagonal
\begin{equation}
\hat{M}^\nu=V^{\nu T}M^\nu V^\nu,\quad
\hat{M}^l= V^{l}_L M^l V^{l\dagger}_ R
\end{equation}
and $M^\nu$ and $M^l$ are given in Eq.~(\ref{leptonmasses}) without the hat.
notice also that 
\begin{equation}
(M^\nu)M^{\nu\dagger}=V^\nu(\hat{M}^\nu)^2V^{\nu \dagger},\quad
M^l(M^l)^\dagger= V^{l\dagger}_L(\hat{M}^l)^2V^l_L,\quad M^{l\dagger}(M^l)= V^{l^\dagger}_R(\hat{M}^l)^2V^l_R
\end{equation}
and that the PMNS matrix is defined as
\begin{equation}
V_{PMNS}=V^{l\dagger}_L V^\nu.
\end{equation}
The matrices $V^\nu,V^l_L,V^l_R$ will appear in some interactions and, for this reason, none of them can be transformed away as it happens in the SM.

\begin{equation}
l^\prime_{L,R}=V^l_{L,R}l_{L,R},\quad \nu^\prime_L=V^\nu \nu_L
\label{sm}
\end{equation}

primed fields denote symmetry eigenstates, unprimed fields are mass eigenstates.

\item  {\bf Auxiliar basis: $G^S$ diagonal:}\\
This basis is defined where $\hat{G}^S$ is diagonal. We denote this basis by a subscript $2$. As $\hat{G}^S$ is symmetric, it can be accomplished by a auxiliar matrix $A$ that obeys,
\begin{equation}
A^TG^S A=G^S_2={\rm Diag}[g_1,g_2,g_3]
\end{equation}
The mass matrix in this basis are related to the general basis as
\begin{equation}
(M^\nu_2)=A^T(M^\nu)A,\quad
M^l_2(M^l_2)^\dagger=A^TM^l(M^l)^\dagger A^*
\end{equation}
while in the diagonal basis as,
\begin{equation}
M^\nu_2=U^*_2(M^\nu)V^{l\dagger}_{2L},\quad
M^l_2(M^l_2)^\dagger=V_{L2}^\dagger(\hat{M}^l)^2V_{L2}
\end{equation}
Where $U_2=A^\dagger V^\nu$ and $V_{L2}^\dagger=A^T V^{l\dagger}_L$. Thus, one can find the PMNS matrix in this basis by
\begin{equation}\label{eq:PMNS_2}
V_{PMNS}=A^*V_{L2}^\dagger AU_2.
\end{equation}
\end{enumerate}

\newpage

\end{document}